\documentclass[letterpaper, 10 pt, conference]{ieeeconf}  % Comment this line out if you need a4paper

\IEEEoverridecommandlockouts                              % This command is only needed if 
\usepackage{graphicx}
\usepackage{url}  %IVM added this because of the URLs in the reference list

\usepackage{amsmath}

\title{\LARGE \bf
A Cough-based deep learning framework for detecting COVID-19
}

\author{Truong~Hoang$^{1}$,
             Lam~Pham$^{2}$,
             Dat~Ngo$^{3}$,
             Hoang D.~Nguyen$^{4}$
\thanks{T. Hoang is with AI Center, FPT Software Company Limited, Vietnam.}%
\thanks{L. Pham is with Competence Unit Data Science \& Artificial Intelligence, Center for Digital Safety \& Security, Austrian Institute of Technology, Austria.}%
\thanks{D. Ngo is with School of Computer Science and Electronic Engineering, University of Essex, UK.}
\thanks{H. D. Nguyen is with School of Computing Science, University of Glasgow, Singapore.}
}

\begin{document}

\maketitle
\thispagestyle{empty}
\pagestyle{empty}

%%%%%%%%%%%%%%%%%%%%%%%%%%%%%%%%%%%%%%%%%%%%%%%%%%%%%%%%%%%%%%%%%%%%
\begin{abstract}

This paper presents a deep learning framework for detecting COVID-19 positive subjects from their cough sounds.
In particular, the proposed approach comprises two main steps. In the first step, we generate a feature representing the cough sound by combining an embedding extracted from a pre-trained model and handcrafted features extracted from draw audio recording, referred to as the front-end feature extraction. Then, the combined features are fed into different back-end classification models for detecting COVID-19 positive subjects in the second step.
Our experiments on the Track-2 dataset of the Second 2021 DiCOVA Challenge achieved the second top ranking with an AUC score of 81.21 and the top F1 score of 53.21 on a Blind Test set, improving the challenge baseline by 8.43\% and 23.4\% respectively and showing deployability, robustness and competitiveness with the state-of-the-art systems.
\newline

\indent \textit{Clinical relevance}--- COVID-19, deep learning, feature extraction, embedding, handcrafted feature. %Convolutional neural network (CNN), recurrent neural network (RNN).
\end{abstract}
%%%%%%%%%%%%%%%%%%%%%%%%%%%%%%%%%%%%%%%%%%%%%%%%%%%%%%%%%%%%%%%%%%%%%%%%%%%%%%%%
\section{INTRODUCTION}

% Theme: Biomedical Engineering transforming the provision of healthcare: promoting wellness through personalized & predictable provision at the point of care

The COVID-19 pandemic has deeply impacted the global health systems with a rising number of 231 million cases and a high death toll of 4.7 million \cite{who}. It is now spanning across 200 countries quickly, and the number of COVID-19 infections per day is consistently reported at an alarming rate without a sign of going down. Therefore, effective solutions for COVID-19 testing on a massive scale is vital for control and mitigate the enormous impacts of the current epidemic. Indeed, if COVID-19 positive subjects can be early detected, it is very useful for self-observation, isolation, and effective treatment methods.

The use of rapid antigen test (ART) and polymerase chain reaction (PCR) tests in popularity has been proven as effective, however, costly and time consuming. With advancement of artificial intelligence, it is promising to alleviate the burden of health care systems through predictable provision at hand for the population. As a result, DiCOVA Challenges are designed to find scientific and engineering insights to the question - Can COVID-19 be detected from the cough, breathing, or speech sound signals of an individual? In particular, while the First 2021 DiCOVA Challenge~\cite{DiCOVA2021} provides a dataset of cough sound only, the Second 2021 DiCOVA Challenge~\cite{seconddicova} provides different sound signals of cough, speech, and breathing.
The audio recordings are gathered from both COVID-19 positive and COVID-19 negative individuals. 
Given the cough, speech, and breath recordings, research community can propose systems for COVID-19 detection, which is potentially deployable on edge devices.% as a COVID-19 screening solution.
%---------------------------------
\begin{figure*}[ht]
    	%\vspace{-0.2cm}
    \centering
    \includegraphics[width =0.9\linewidth]{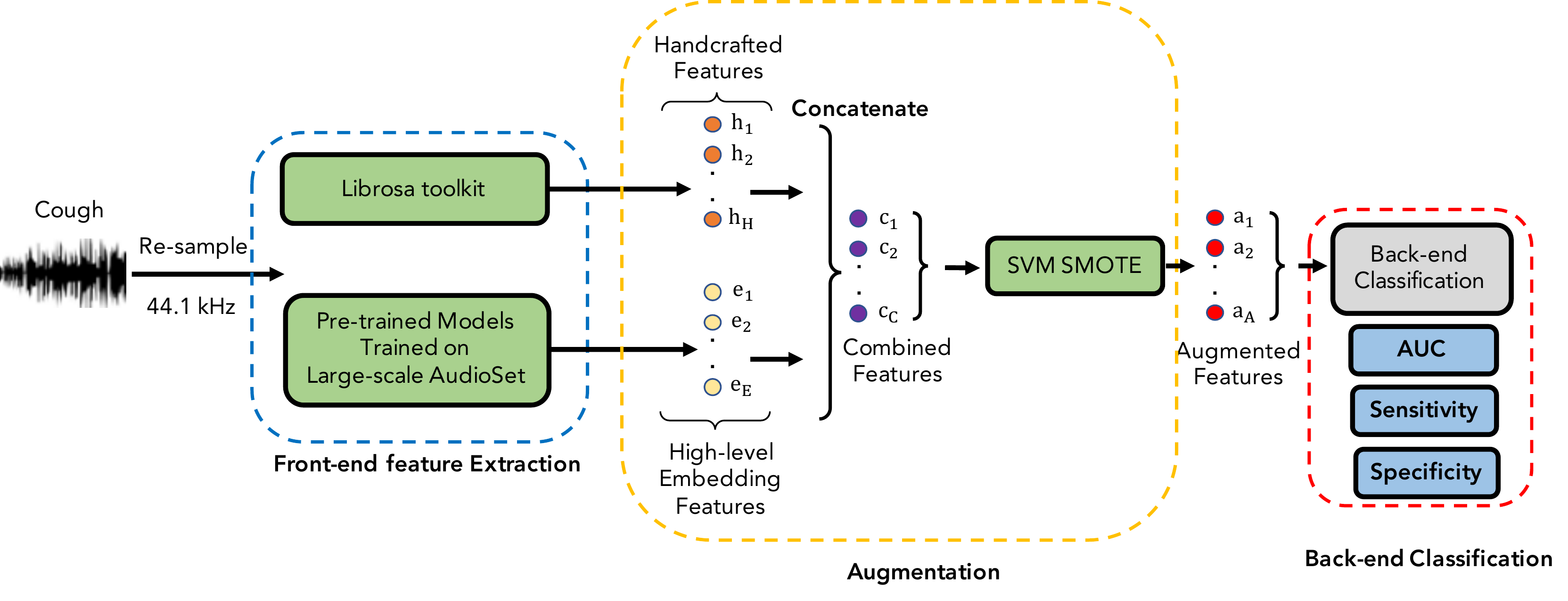}
    	%\vspace{-0.5cm}
	\caption{The high-level architecture of deep learning framework proposed.}
    \label{fig:A1}
            %	\vspace{-0.3cm}
\end{figure*}

There have been multiple studies~\cite{dicova_2021_002, dicova_2021_003} gathering insights on the possibility of acoustics based COVID-19 diagnosis. 
Focusing on the cough sound, recent researchers show that it potential to detect COVID-19 through evaluating coughing. For an example, a machine learning-based framework proposed in~\cite{dicova_2021_03} utilized handcrafted features and Support Vector Machine (SVM) model, achieved the AUC score of 85.02 on the First DiCOVA dataset~\cite{DiCOVA2021}. 
Further exploration on this dataset proposed a deep learning framework in~\cite{dicova_2021_00}, which use the ConvNet model incorporated with Data Augmentation, achieved the best AUC score of 87.07 and presented the top-1 position in the First DiCOVA Challenge.
Focusing on feature extraction, Madhu et al.~\cite{dicova_2021_01} combined the Mel-frequency cepstral coefficients (MFCC) with the delta features (i.e. The delta features are extracted from a complicated framework using Long Short-Term Memory (LSTM), Gabor filter bank, and the Teager energy operator (TEO) in the order). By using the combined features and the back-end LightGBM model, the authors can achieve the AUC score of 76.31 with the First DiCOVA dataset~\cite{DiCOVA2021}.
Similarly, Vincent et al.~\cite{dicova_2021_02} conducted extensive experiments to evaluate the role of the feature extraction.
In particular, they proposed to use three types of features: (1) Handcrafted features extracted by openSMILE toolkit~\cite{opensmile}, (2) the deep features extracted from different pre-trained VGGish networks which are trained with AudioSet~\cite{audioset}, and (3) the deep features extracted from different standard pre-trained models (ResNet50, DenseNet121, MobileNetV1, etc.) trained with Imagenet dataset. They then obtained the best AUC score of 72.8 on the First 2021 DiCOVA dataset~\cite{DiCOVA2021} by using the deep features extracted from the pre-trained VGG16 (i.e. The pre-trained VGG16 was trained with AudioSet) and the back-end LSTM-based classification.
Recently, a benchmark dataset of cough sound for detecting COVID-19~\cite{coughcelldetect, diagnosis},  which was recorded on mobile phone, has been published. 
%Notably, the current achievement of 98\% accuracy on this dataset shows potential to apply as an effective solution of COVID-19 testing.

In this paper, we also aim to explore cough sounds, then propose a framework for detecting COVID-19.
We mainly contribute: (1) discriminative features by combining handcrafted feature and embedding based feature for COVID detection by analysing cough sound,
%COVID-19 domains might be new; therefore, further exploration of existing ways will also help to advance this research field.
%By conducting extensive experiments, we indicate that a combination of handcrafted feature and embedding-based feature is effective to representing cough sound input, 
and (2) a robust framework which can be further developed on edge devices for an application of COVID-19 testing. 
Our experiments are conducted on the Track-2 dataset in the Second 2021 DiCOVA Challenge (i.e. only contains cough sounds).

%The remaining of our paper is organized as follows: Section \ref{sec:dataset} presents the Second 2021 DiCOVA Challenge as well as the Track-2 dataset, evaluation setting, and metrics. Section \ref{sec:method} presents the proposed deep learning framework. Next, Section \ref{sec:result} presents and analyses the experimental results. Finally, Section \ref{sec:conclusion} presents the conclusion and future work.

%\begin{figure}[t]
%	\centering
%    \includegraphics[width=3.3in]{waveform.png}
%    \caption{The waveform of the Cough, Breathing, Speech sound from the Second 2021 DiCOVA Challenge~\cite{seconddicova}.}
%    \label{fig:waveform}
%\end{figure}
%---------------------------------

\section{The Track-2 dataset of cough sounds in The Second 2021 DiCOVA Challenge}
\label{sec:dataset}
The Second 2021 DiCOVA Challenge uses a subset of the Coswara dataset \cite{seconddicova} collected between April 2020 and July 2021 from the age group of 15 to 90. 
The challenge provides a dataset of different sound signals: cough, speech, and breathing gathered from both COVID-19 positive and non-COVID-19 individuals. 
%as shown in Fig. \ref{fig:waveform}.
Given cough, speech, and breath sounds, the Second 2021 DiCOVA Challenge proposes four tracks that aim to detect COVID-19 positive subjects by exploring only breathing (Track-1), only cough (Track-2), only speech (Track-3), or all sound signals (Track-4).

As we aim to focus on cough sounds, which is also the First 2021 DICOVA Challenge~\cite{DiCOVA2021}, only Track-2 dataset in the Second 2021 DiCOVA is explored in this paper. 
The Track-2 dataset provides a Development set of 965 audio recordings and a Blind Test set of 471 audio recordings.
All audio recordings of cough sounds are not less than 500 milliseconds and recorded with different sample rates.
While Development set is used for training, and then obtaining the best model, Blind Test set is used for evaluating and comparing the systems' performance submitted.
In Development set, there are a total of 793 negative labels and 172 positive labels, which shows a significantly imbalanced dataset~\cite{imbalanced}. 

The `Area under the ROC curve' (AUC) is used as the primary evaluation metric in the Second 2021 DiCOVA Challenge. 
%The curve is obtained by varying the decision threshold between 0 and 1 with a step size of 0.0001. 
%Additionally, the Specificity (Spec.) and Sensitivity (Sen.) are obtained at every threshold value.
Additionally, the Sensitivity (Sens.) and the Specificity (Spec.)
%, which are computed at certain threshold value, 
are used as the secondary evaluation metrics (Note that Spec. is required to be equal or greater than 95\%).
%greater than or equal to 95\% is used as the secondary evaluation metrics.
The Leaderboard provides the evaluation of the submitted systems on Blind Test set as well as the average performance on five-fold cross validation from Development set (Avg. AUC) \cite{seconddicova}. 
%The score files that participants are required to submit show the probability of COVID positive for each test recording \cite{seconddicova}.

\section{Deep Learning Framework proposed}
\label{sec:method}

\subsection{High-level architecture of deep learning framework}

The overall framework architecture is described as Fig.~\ref{fig:A1}. As the audio recordings show different sample rates, they are firstly re-sampled to 44.1 kHz using a mono channel. Then, the re-sampled recordings are inputted into the front-end feature extraction, where embedding-based features and handcrafted features are extracted and concatenated to obtain the combined features. To deal with the issue of imbalanced dataset mentioned in Section \ref{sec:dataset}, SVM-based SMOTE method, a variant of SMOTE algorithm which uses SVM algorithm to detect neighbor samples~\cite{smote}, is applied on the combined features to make sure the equal number of positive and negative samples. Finally, the combined features after augmentation are fed into different back-end classification models for detecting COVID-19 positive cases.

\subsection{The front-end feature extraction}

In this step, we propose a method to create combined features by combining handcrafted features and embedding-based features extracted from pre-trained models for COVID detection.
Regarding handcrafted features, 64 Mel-frequency cepstral coefficients (MFCCs), 12 Chromatic (Chroma), 128 Mel Spectrogram (Mel), 1 Zero-Crossing rate, 1 Gender, and 1 Duration are utilized in this paper. 
These handcrafted features are used as they are popularly adopted in speech processing and show robustness in the First 2021 DiCOVA Challenge~\cite{dicova_2021_01, dicova_2021_02, dicova_2021_03}. 
To extract these handcrafted features, Librosa Toolkits \cite{mcfee2015librosa}, a powerful library of audio signal processing, is used in this paper with the window size, FFT number, hop size set to 2048, 2048, 512.

As regards the embedding features, we evaluate different embedding features which are extracted from a wide range of pre-trained models: YAMNet~\cite{yamnet}, Wave2Vec~\cite{wav2vec}, TRILL \cite{trill}, and the COMPARE 2016 feature sets \cite{compare2016} using OpenSMILE~\cite{opensmile} toolkit.
As using these pre-trained models shows effective for a wide range of down-stream classification tasks (i.e. The pre-trained TRILL model with AudioSet~\cite{audioset} proves robust for various classification tasks on non-semantic speech signal such as speaker identity, language, and emotional state in~\cite{trill}), these embeddings are expected to work well with the Track-2 dataset of cough sounds in the Second 2021 DiCOVA Challenge.
By using the pre-trained models, when we feed a cough recording into the pre-trained models, 2-dimensional embeddings are extracted. 
We then compute mean and standard deviation across the time dimension, concatenating mean and standard deviation to obtain one 1-dimensional embedding which represents for one input audio sample. 
The embedding-based feature is then concatenated with the handcrafted feature mentioned above to create a combined feature.
Finally, the combined feature is scaled into a range of [0:1] before conducting data augmentation and fed into the back-end classification models.

\subsection{The back-end classification models}

In this paper, we evaluate different back-end classification models: Light Gradient Boosting Machine (LightGBM),  Random Forrest (RF), Support Vector Machine (SVM), Multi-layer Perceptron (MLP), and Extra Tree Classifier (ETC).
To obtain the hyper-parameters used for optimizing these back-end classifiers, we apply the Grid Search algorithm from the Optuna framework~\cite{optuna_2019}. Settings of these back-end classification models are described in Table~\ref{table:set_model} and all these models are implemented by using Scikit-Learn toolkit~\cite{scikit-learn}.
%---------------------
\begin{table}[t]
    \caption{Back-end classification models and setting parameters.} 
    \centering
    \scalebox{0.95}{
    \begin{tabular}{|l |l|} 
       \hline
     \textbf{Models} & \textbf{Setting Parameters}  \\
    \hline
    \hline
    Support Vector Machine (SVM) &  C=1.0 \\
        & Kernel=`RBF'   \\
        & gamma=`scale'   \\
            \hline
    Random Forest (RF) & Max Depth of Tree = 20, \\
                   & Number of Trees = 100  \\
                       \hline
      & Two hidden layer (4096 nodes),  \\
    Multilayer Perceptron (MLP)     & Adam optimization, \\
         %& Hidden layer sizes = (2048, 2048), \\
         & Max iter = 200 \\
         & Learning rate = 0.001, \\
         & Entropy Loss \\
   \hline
    ExtraTreesClassifier (ETC) & Max Depth of Tree = 20 \\
    \hline
     & learning rate = 0.03 \\
      LightGBM~\cite{LightGBM}
             %& boosting type = `gbdt', \\
             & objective = `binary' \\
             %& tree\_learner = `serial' \\
             & metric = `auc' \\
             & subsample = 0.68 \\
             & colsample\_bytree = 0.28 \\
             & early\_stopping\_rounds = 100 \\
             & num\_iterations = 10000 \\
             & subsample\_freq = 1 \\
             %& reg\_lambda = 2 \\
             %& reg\_alpha = 1 \\
             %& num\_leaves = 500 + seed * 25 \\
             %& random\_state = seed \\
    \hline
    \end{tabular}
    %\vspace{-0.5cm}
    }
    \label{table:set_model} 
\end{table}
To obtain results, each classification model is run with 10 seeds numbered from 0 to 9. The output of the cross-validation session will calculated by using an average of 10 seeds.
%As the seed value equals the sequence number multiples of 25, the $num\_leaves$ setting has the values of 500, in turn, 525, ..., 700, 725 across 10 seeds. 
%Then, soft voting between seeds is used to obtain the final result, which is the average of 10 seeds. 
The GTX 1080 Titan GPU environment is used for running classification experiments. 

\section{Experimental results and discussion}
\label{sec:result}

\subsection{Performance comparison across different features}

To evaluate different features extracted, we keep the back-end classification model of LightGBM unchanged while replacing different input features: Handcrafted feature, YAMNet based embedding, COMPARE 2016 based embedding, Wave2Vec based embedding, TRILL based embedding, handcrafted \& YAMNet, handcrafted \& COMPARE 2016, handcrafted \& Wave2Vec, and handcrafted \& TRILL features.
As the results are shown in Table~\ref{table:feature}, it can be seen that TRILL-based embedding outperforms the other single features, reporting an Avg. AUC score of 73.77 and 80.57 on Development set and Blind Test set, respectively.

When we combine the handcrafted feature with different embedding-based features of YAMNet, COMPARE 2016, and TRILL, it is effective to improve the performance, reporting Avg. AUC scores of 77.33, 77.19, and 77.18, respectively compared with 72.62 of using handcrafted feature only. 
The best performance is obtained from the combination of the handcrafted feature and TRILL-based embedding feature, achieving the AUC, Sens., and Spec. scores of 81.21, 48.33, and 95.13 respectively on Blind Test set. 
%\begin{itemize}
%    \item CI of Fold 0: 0.77591554357 (0.76613990697, 0.78569118018) AUC
%    \item CI of Fold 1: 0.78206109613 (0.77020047524, 0.79392171703) AUC
%    \item CI of Fold 2: 0.74490026954 (0.73371756393, 0.75608297514) AUC
%    \item CI of Fold 3: 0.77251350351 (0.76465979147, 0.78036721555) AUC
%    \item CI of Fold 4: 0.783701602 (0.77185831239, 0.79554489162) AUC
%\end{itemize}

\subsection{Performance comparison across different classification models}
As we obtained the best handcrafted \& TRILL-based embedding feature from the experiments above, we now evaluate how back-end classification models affect the performance.
To this end, we keep the handcrafted \& TRILL-based embedding feature unchanged while replacing LightGBM by different back-end classification models of Support Vector Machine (SVM), Random Forest (RF), Extra Trees Classifier (ETC), and Multi-layer perceptron (MLP). 
As the results are shown in Table~\ref{table:model}, the LightGBM model still achieves the best scores. 
Meanwhile, the other models show competitive results, reporting Avg. AUC/Blind Test AUC scores of 75.54/76.27, 74.04/78.72, 72.50/76.34, and 74.87/77.51 for SVM, RF, MLP, and ETC respectively. 
%---------------------
\begin{table}[t]
    \caption{Performance comparison across different features with the back-end LightGBM model.} %(the best performance results are in \textbf{bold}).} 
        	%\vspace{0.2cm}
    \centering
    \scalebox{0.75}{
    \begin{tabular}{|l |c| c| c| c|} 
       \hline
     \textbf{Extracted Features} & \textbf{AUC} &  \textbf{Sens.} &  \textbf{Spec.} & \textbf{Avg. AUC} \\
                   & \textbf{(Blind Test)} & \textbf{(Blind Test)} & \textbf{(Blind Test)} & \textbf{(Development)} \\
    \hline
    \hline
    Handcraft & 76.36 & 36.66 & 95.13 & 72.62  \\
    YAMNet~\cite{yamnet} & 67.24 & 21.51 & 95.13 & 67.31  \\
    COMPARE 2016~\cite{compare2016} & 63.18 & 15.00 & 95.13 & 71.00  \\
    Wave2Vec~\cite{wav2vec} & 58.86 & 06.66 & 95.13 & 58.75  \\
    TRILL~\cite{trill} & 80.57 & 43.33 & 95.13 & 73.77 \\
    Handcraft + YAMNet & 77.27 & 41.67 & 95.13 & 77.33  \\
    Handcraft + COMPARE 2016 & 69.14 & 25.00 & 95.13 & 77.19  \\
    Handcraft + Wave2Vec & 71.00 & 25.00 & 95.13 & 71.47  \\
    %Handcraft + TRILL + Delta-Delta MFCC & 80.05 & 45.00 & 95.13 & 75.31  \\
    \textbf{Handcraft + TRILL} & \textbf{81.21}  & \textbf{48.33} & \textbf{95.13} & \textbf{77.18}\\
    \hline
    \end{tabular}
    }
    \label{table:feature} 
\end{table}
\begin{table}[t]
    \caption{Performance comparison across different back-end classification models with handcrafted and TRILL based embedding features} %(the best performance results are in \textbf{bold}).} 
        	%\vspace{0.2cm}
    \centering
    \scalebox{0.8}{
    \begin{tabular}{|l |c| c| c| c|} 
       \hline
     \textbf{Back-end} & \textbf{AUC} &  \textbf{Sens.} &  \textbf{Spec.} & \textbf{Avg. AUC} \\
      \textbf{Classification}             & \textbf{(Blind Test)} & \textbf{(Blind Test)} & \textbf{(Blind Test)} & \textbf{(Development)} \\
    \hline
    \hline
    SVM & 76.27 & 36.66 & 95.13 & 75.54  \\
    RandomForest & 78.72 & 36.66 & 95.13 & 74.04  \\
    Multi-layer Perceptron & 76.34 & 31.66 & 95.13 & 72.50  \\
    ExtraTreesClassifier & 77.51 & 38.33 & 95.13 & 74.87  \\
    \textbf{LightGBM} & \textbf{81.21}  & \textbf{48.33} & \textbf{95.13} & \textbf{77.18}\\
    \hline
    \end{tabular}
    }
    \vspace{-0.2cm}
    \label{table:model} 
\end{table}
%---------------------

To make sure the best score is from the combination of the handcrafted feature and TRILL-based embedding feature for the front-end feature extraction and LightGBM for back-end classification, we conducted 10 times of running the experiments on five folds of Development set.
We achieve an average confidence interval (CI) of (0.76610763001, 0.77752917589), which matches the AUC score of 77.18 from Table~\ref{table:model}.

\subsection{Performance comparison across the top-10 systems submitted for the Track-2 dataset of the Second 2021 DiCOVA Challenge}
To compare the state-of-the-art systems, we joined in Track-2 of the Second 2021 DiCOVA Challenge, submitted our proposed system, and compare with the other submissions. 
The Table~\ref{table:sota} presents the performance comparison across the top-10 systems submitted for the Track-2 of the Second 2021 DiCOVA Challenge.
As shown in Table~\ref{table:sota}, our best results from handcrafted \& TRILL-based embedding features and LightGBM model achieved the top-2 AUC score of 81.21, only after the top-1 AUC score of 81.97.
Nevertheless, we achieved a sensitivity score of 48.33, precision of 59.18 and F1 score of 53.21, which has significant improvements of 31.8\%, 12.99\% and 23.4\% respectively from both baseline and the top-1 submission. Despite the sense of high AUC, it is important to to evaluate the suitably calibrated probabilities due to the nature of COVID-19 screening and imbalanced dataset.
%, which significantly outperforms the other systems (i.e. The top-1 AUC system only achieved a sensitivity score of 36.67).
%As we can achieve the highest Sens. score, our Recision and F1 scores of 59.18 and 53.21 respectively are also the top-1 compared with the other systems. 
%It can be seen that our Spec. result on the blind Test set is very competitive to the top-1 and top-2 systems. 
These results demonstrate that our proposed system is deployable, robust, competitive, and has the potential to be further applied on edge devices for detecting COVID-19.

%---------------------
\begin{table}[t]
    \caption{Performance comparison on Blind Test set across the top-10 systems submitted and the challenge baseline.} 
        	%\vspace{0.2cm}
    \centering
    \scalebox{0.9}{
    \begin{tabular}{|l |c| c| c| c| c| } 
        \hline   
        %\textbf{Systems} & \textbf{AUC} &  \textbf{Sens.} &  \textbf{Spec.} &  \textbf{Prec.} & \textbf{F1 Score} \\
                   %& \textbf{(Blind Test)} & \textbf{(Blind Test)} & \textbf{(Blind Test)} & \textbf{(Blind Test)} \\
         \textbf{Systems} & \textbf{AUC} &  \textbf{Sens.} &  \textbf{Spec.} & \textbf{Prec.} & \textbf{F1 Score} \\

    \hline
    \hline
   1st system & \textbf{81.97} & 36.67 & 95.13 & 52.38 & 43.14  \\
   \textbf{2nd (Our system)} & 81.21  & \textbf{48.33} & 95.13 & \textbf{59.18} & \textbf{53.21}  \\
    3rd system & 80.12 & 35.00 & 95.13 & 51.22 & 41.58  \\
    4th system & 79.06 & 35.00 & 95.13 & 51.22 & 41.58  \\
    5th system & 77.85 & 46.67 & 95.13 & 58.33 & 51.85  \\
    6th system & 77.60 & 33.33 & 95.13 & 50.00 & 40.00  \\
    7th system & 76.98 & 40.00 & 95.13 & 54.55 & 46.15  \\
    8th system & 76.36 & 30.00 & 95.13 & 47.37 & 36.73  \\
    9th system & 75.95 & 40.00 & 95.13 & 54.55 & 46.15  \\
    10th system & 75.71 & 35.00 & 95.13 & 51.22 & 41.58  \\
        \hline
    \textbf{Challenge baseline} & 74.89 & 36.67 & 95.13 & 52.38 & 43.14  \\
    \hline 
    \end{tabular}
    }
    \vspace{-0.4cm}
    \label{table:sota} 
\end{table}

\section{Conclusion and Future Work}
\label{sec:conclusion}

This paper has presented a robust deep learning framework for detecting COVID-19 positive subjects by exploring cough sound inputs. 
By conducting extensive experiments on the Track-2 of the Second 2021 DiCOVA Challenge, our proposed framework with a discriminative combined feature (handcrafted feature \& embedding based feature from pre-trained TRILL model) and LightGBM model achieved the high performance in a stable manner, showing a potential for a real-life application.

%For handcrafted feature, we approach handcrafted feature due to the unbalanced and small DiCOVA dataset (i.e. Frame-based framework with handcrafted feature and traditional machine learning should be suitable for small datasets). For embedding based feature, pre-trained TRILL \cite{trill} model on the large scale AudioSet presents a large contextwindow of the network and generality of training triplet loss in the pre-trained TRILL broadly preserves acoustic characteristics instead of prematurely focusing on certain aspects. Therefore, embedding feature extracted from the pre-trained TRILL model, which show effective for various down-stream tasks \cite{trill}, should be effective and well-represented for cough sound. COVID-19 domains might be new; thus, further exploration of existing ways will also help to advance this research field. 
Our further research are to deeply analyse roles of input features and focus on different sound representations such as Chroma Feature, Spectral Contrast, Tonnetz, etc \cite{acoustics2021}, as well as to explore breathing, speech sounds provided by the Second 2021 DiCOVA Challenge.


%\begin{thebibliography}{99}
%\bibliographystyle{IEEEbib}
\bibliographystyle{IEEEtran}
\bibliography{paper10_v1}

%\end{thebibliography}
\end{document}